# Tunneling under Coherent Control by Sequences of Unitary Pulses

Rajdeep Saha and Victor S. Batista*

*Yale University, Department of Chemistry, P.O. Box 208107, New Haven, CT 06520-8107*

**Abstract:** A general coherent control scenario to suppress, or accelerate, tunneling of quantum states decaying into a continuum, is investigated. The method is based on deterministic, or stochastic, sequences of unitary pulses that affect the underlying interference phenomena responsible for quantum dynamics, without inducing decoherence, or collapsing the coherent evolution of the system. The influence of control sequences on the ensuing quantum dynamics is analyzed by using perturbation theory to first order in the control pulse fields and compared to dynamical decoupling (DD) protocols and sequences of pulses that collapse the coherent evolution and induce quantum Zeno (QZE) or quantum anti-Zeno effects (AZE). The analysis reveals a subtle interplay between coherent and incoherent phenomena and demonstrating that dynamics analogous to evolution due to QZE or AZE can be generated from stochastic sequences of unitary pulses when averaged over all possible realizations.

## I. INTRODUCTION

Advancing our understanding of coherent control techniques to accelerate, or suppress, tunneling of quantum states decaying into a manifold of continuum states is a problem of great technological interest.[1-3] During the past 30 years, several coherent control methods have been developed and optimized to manipulate a wide range of quantum processes,[4-21] including tunneling dynamics.[13-15][16-19] This paper focuses on one of the most recently proposed methods,[15,22-24] based on sequences of unitary pulses that repetitively change the phases of interfering states, responsible for quantum dynamics, without inducing decoherence, or collapsing the coherent evolution of the system. The method has been numerically demonstrated as applied to control super-exchange electron tunneling dynamics in monolayers of adsorbate molecules functionalizing semiconductor surfaces when using either deterministic, or stochastic, sequences of unitary phase-kick ($2\pi$ pulses).[25-30] However, the underlying control mechanism induced by sequences of phase-kick pulses has been difficult to elucidate from a cursory examination of the ensuing dynamics. This paper reports a rigorous theoretical analysis of the origin of quantum control as resulting from the interplay between coherent and incoherent phenomena induced by deterministic, or stochastic, sequences of pulses. Control is analyzed by perturbation theory to first order in the pulse fields and compared to dynamical decoupling (DD) protocols[31-33] and sequences of pulses that periodically collapse the coherent evolution[34-37] yielding to dynamics modulated by quantum Zeno (QZE) and quantum anti-Zeno (AZE) effects.[15,22,38] The reported results provide fundamental insights on the origin of suppression of quantum tunneling by sufficiently frequent perturbation pulses, and acceleration induced by pulses separated by finite time intervals. The analytic expressions reported for the description of short-time dynamics also provide understanding on the effect of randomization of pulse sequences and clarify how the ensuing dynamics depends on the average time-period between perturbational phase-kick pulses when averaged of all possible realizations of control sequences. This results are particularly valuable since stochastic pulse sequences have already been demonstrated to achieve control in condensed material systems,[29,30] or predicted to outperform deterministic pulsed schemes in control of quantum coherences.[15,22] Considering that current laser technology can produce a wide range of pulses with ultrashort time resolution and extremely high-peak power, it is natural to expect that the quantum control techniques explored in this paper should raise significant experimental interest.[34,39]

The paper is organized as follows. Section II introduces the system and the description of spontaneous decay due to tunneling into a continuum. Section III introduces coherent control based on equally time-spaced phase-kick pulses, as applied to the acceleration, or suppression, of tunneling into a continuum. Section IV analyzes a generalization of the method to sequences of randomly time-spaced pulses. Section V explores stochastic sequences, averaged over all possible realizations, as compared to DD protocols, and the inherent similarities with Quantum Zeno and Anti-Zeno effects. Concluding remarks and future directions are presented in Section VI.

## II. TUNNELING INTO A CONTINUUM

We consider the system, depicted in Fig. 1, initially prepared in a bound state $|s\rangle$ coupled to a continuum, as described by the following Hamiltonian:[39,40]

$$\hat{H} = \omega_s |s\rangle\langle s| + \sum_k \omega_k |k\rangle\langle k| + \sum_k \left( V_{ks} |k\rangle\langle s| + V_{sk} |s\rangle\langle k| \right), \quad (1)$$

where $|s\rangle$ and $|k\rangle$ are stationary eigenstates of $\hat{H}$, when $V_{ks} = 0$, with energies $\omega_s$ and $\omega_k$, respectively. For simplicity, notation is kept in atomic units (with $\hbar = 1$). When $V_{ks} \neq 0$, state $|s\rangle$ is non-stationary. Therefore, a system initially prepared in state $|s\rangle$ spontaneously decays by tunneling into the continuum. In the absence of external perturbations, the time-evolution is described by the time-dependent wavefunction,

$$\Psi(t) = \alpha_s(t)e^{-i\omega_s t}|s\rangle + \sum_k \beta_k(t)e^{-i\omega_k t}|k\rangle, \quad (2)$$

with $\alpha_s(0) = 1$, and $\beta_k(0) = 0$ for all $|k\rangle$.

The equations of motion of the time-dependent expansion coefficients, introduced by Eq. (2), are obtained by solving the time-dependent Schrödinger equation, as follows:

$$\dot{\alpha}_s = -i \sum_k V_{sk} e^{i(\omega_s - \omega_k)t} \beta_k, \quad (3)$$

$$\dot{\beta}_k = -i V_{ks} e^{i(\omega_k - \omega_s)t} \alpha_s \quad (4)$$

Integrating Eq. (4) from time $t_b$ to time $t$, yields:

$$\beta_k(t) - \beta_k(t_b) = -i \int_{t_b}^{t} V_{ks} e^{i(\omega_k - \omega_s)t'} \alpha_s(t')\, dt', \tag{5}$$

and substituting Eq. (5) into Eq. (3), gives:

$$\dot{\alpha}_s(t) = -\int_{t_b}^{t} \sum_k |V_{ks}|^2 e^{i(\omega_s - \omega_k)(t-t')} \alpha_s(t')\, dt' - i\sum_k V_{sk} \beta_k(t_b). \tag{6}$$

Equation (6) can be solved exactly by using standard Laplace transform techniques.[34,41] However, for short times, the solution of Eq. (0.6) can be approximated as follows:[41,42]

$$\alpha_s(t) \approx \alpha_s(t_b)\left(1 - \sum_k |V_{ks}|^2 \int_{t_b}^{t}(t-t')e^{i(\omega_s-\omega_k)(t'-t_b)}dt'\right) - i\sum_k V_{sk} \int_{t_b}^{t} e^{i(\omega_s-\omega_k)t'}\beta_k(t_b)dt', \tag{7}$$

where we introduced the approximation $\alpha_s(t') = \alpha_s(t_b)$, as shown in Appendix A. Similarly, the expansion coefficients for states $|k\rangle$ from Eq. (5) gives:

$$\beta_k(t) \approx \beta_k(t_b) - i\alpha_s(t_b) V_{ks} \int_{t_b}^{t} e^{i(\omega_k - \omega_s)t'}\, dt'. \tag{8}$$

Equation (7) yields the standard expression for the spontaneous population decay of state $|s\rangle$, due to coupling to the manifold of continuum states $|k\rangle$, as follows:[15,22,43]

$$P_s(t) = |\alpha(t)|^2 = 1 - \sum_k \frac{|V_{ks}|^2}{\left(\frac{\omega_s - \omega_k}{2}\right)^2} \sin^2\left((\omega_s - \omega_k)\frac{t}{2}\right). \tag{9}$$

Sections III–V show that the spontaneous, decay described by Eq. (9), can be suppressed, or accelerated, by perturbing the system with a train of pulses (Fig. 1) that change the *phase* of the wavefunction component along state $|s\rangle$, relative to the other terms in the coherent state expansion of Eq. (2). Sec. V also shows that Eq. (9) is recovered in the limit where the pulses have a low probability of inducing changes of phase.

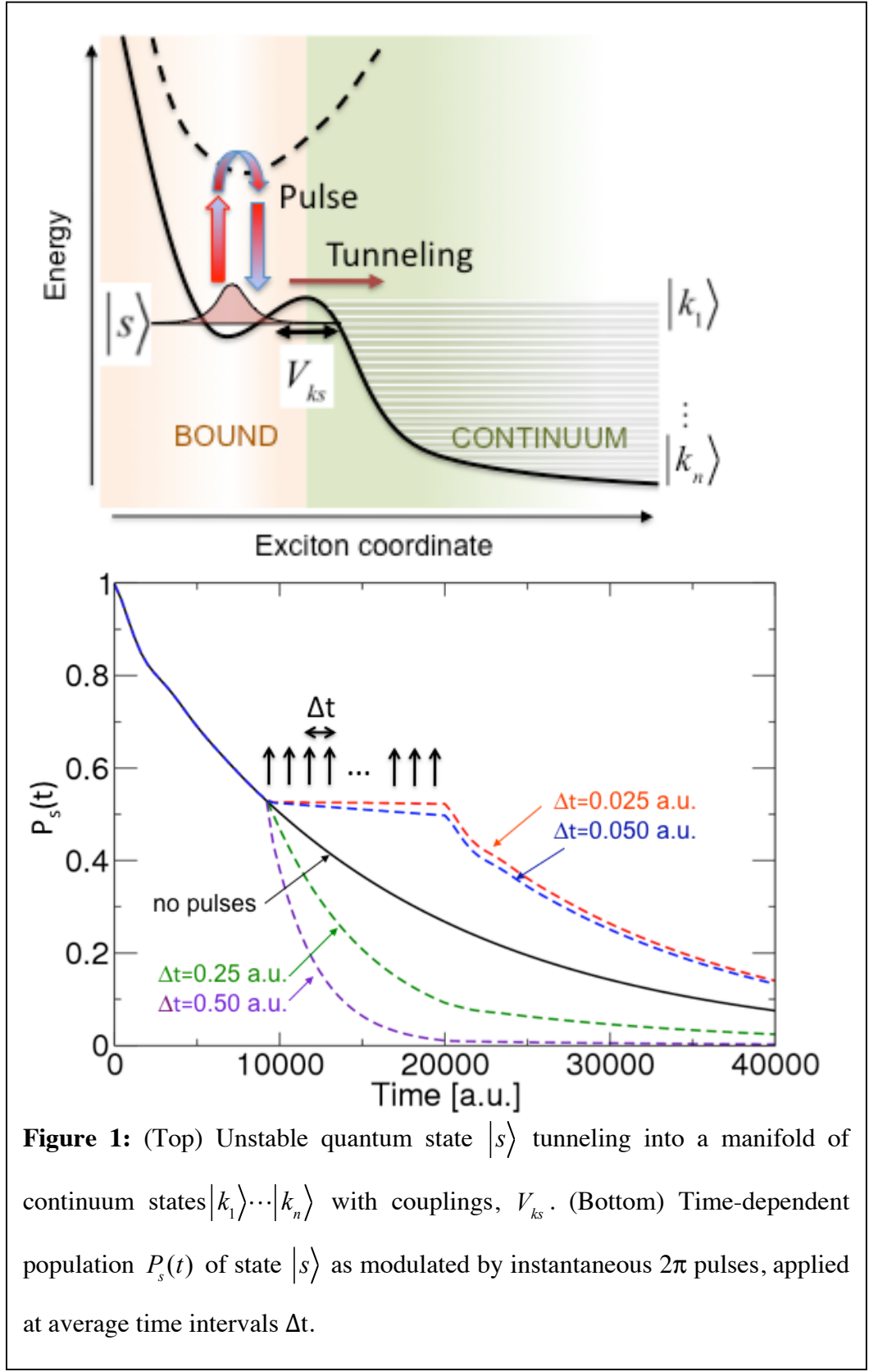

**Figure 1:** (Top) Unstable quantum state $|s\rangle$ tunneling into a manifold of continuum states $|k_1\rangle \cdots |k_n\rangle$ with couplings, $V_{ks}$. (Bottom) Time-dependent population $P_s(t)$ of state $|s\rangle$ as modulated by instantaneous $2\pi$ pulses, applied at average time intervals $\Delta t$.

## III. PERIODIC PULSING

Consider the evolution of the system, introduced in Sec. II, as perturbed by two consecutive instantaneous pulses $\hat{Q}$, spaced by a time-interval $\Delta t$, as follows:

(1) Evolve the system for a short time period, $\Delta t$, using Eqs. (7) and (8).

(2) Apply an instantaneous pulse, $\hat{Q}$.

(3) Continue the evolution, from $t = \Delta t$ to $t = 2\Delta t$, according to Eqs. (7) and (8).

(4) Apply another pulse, $\hat{Q}$.

Repeating steps 1–4, $n$ times, evolves the system to time $t = 2n\Delta t$, yielding the expansion coefficients $\alpha_s(2n\Delta t)$ and $\beta_k(2n\Delta t)$ for states $|s\rangle$ and $|k\rangle$, respectively.

For the specific case of sequences of $2\pi$ pulses, each pulse $\hat{Q}$ changes the sign of the projection of the time-evolved wavefunction along the direction of $|s\rangle$, as follows:[15,22,43]

$$\begin{aligned}\hat{Q}|\psi\rangle &= |\psi\rangle - 2|s\rangle\langle s|\psi\rangle \\ &= \sum_k |k\rangle\langle k|\psi\rangle - |s\rangle\langle s|\psi\rangle\end{aligned} \quad (10)$$

leaving unaffected the projection of $|\psi\rangle$ along the manifold of states $|k\rangle$ in the continuum. Therefore, $2\pi$ pulses can be represented as $Q_s = 1 - |s\rangle\langle s|$,[23,24,27,29,44] yielding the following evolution for the expansion coefficients:

$$\beta_k(\Delta t) = \beta_k(0) - iV_{ks}\left(\frac{e^{i(\omega_k-\omega_s)\Delta t}-1}{i(\omega_k-\omega_s)}\right)\alpha_s(0)$$

$$\alpha_s(\Delta t) = \alpha_s(0)\left(1 - \int_0^{\Delta t}(\Delta t - t')K(t')\,dt'\right) - i\sum_k V_{sk}\int_0^{\Delta t} e^{i(\omega_s-\omega_k)t}\beta_k(0)\,dt$$

$$\alpha'_s(\Delta t) = -\alpha_s(\Delta t)$$

$$\beta_k(2\Delta t) = \beta_k(\Delta t) - iV_{ks}\left(\frac{e^{i(\omega_k-\omega_s)2\Delta t} - e^{i(\omega_k-\omega_s)\Delta t}}{i(\omega_k-\omega_s)}\right)\alpha'_s(\Delta t) \qquad (11)$$

$$\alpha'_s(2\Delta t) = \alpha'_s(\Delta t)\left(1 - \int_{\Delta t}^{2\Delta t}(2\Delta t - t')K(t' - \Delta t)\,dt'\right) - i\sum_k V_{sk}\int_{\Delta t}^{2\Delta t} e^{i(\omega_s-\omega_k)t}\beta_k(\Delta t)\,dt$$

$$\alpha_s(2\Delta t) = -\alpha'_s(2\Delta t)$$

where $K(t) = \sum_k |V_{ks}|^2 e^{i(\omega_s-\omega_k)t}$. Collecting the expressions, introduced by Eq. (11), with $\alpha_s(0) = 1$ and $\beta_k(0) = 0$, we obtain:

$$\begin{aligned}
\alpha_s(2\Delta t) = &\left\{\alpha_s(0)\left(1 - \int_0^{\Delta t}(\Delta t - t')K(t')\,dt'\right) - i\sum_k \int_0^{\Delta t} V_{sk} e^{i(\omega_s-\omega_k)t'}\beta_k(0)\,dt'\right\} \times \\
&\left\{\left(1 - \int_{\Delta t}^{2\Delta t}(2\Delta t - t')K(t' - \Delta t)\,dt'\right)\right\} + i\sum_k \int_{\Delta t}^{2\Delta t} V_{sk} e^{i(\omega_s-\omega_k)t'}\beta_k(\Delta t)\,dt' \\
= &1 - \underbrace{\int_0^{\Delta t}(\Delta t - t')K(t')\,dt' - \int_{\Delta t}^{2\Delta t}(2\Delta t - t')K(t' - \Delta t)\,dt'}_{I} \\
&\underbrace{-i\sum_k \int_0^{\Delta t} V_{sk} e^{i(\omega_s-\omega_k)t'}\beta_k(0)\,dt'}_{II} + \underbrace{i\sum_k \int_{\Delta t}^{2\Delta t} V_{sk} e^{i(\omega_s-\omega_k)t'}\beta_k(\Delta t)\,dt'}_{III}
\end{aligned} \qquad (12)$$

where terms of $O(|V_{ks}|^3)$ have been neglected. Terms II and III introduce the couplings to the continuum with have opposite signs, due to the nature of the pulsing process, as explained in Appendix B.

Repeating the process described above, $n$ times, we obtain $\alpha_s(2n\Delta t)$. The contributions to $\alpha_s(2n\Delta t)$ from term I are:

$$\text{Terms from } I = \left( \sum_k |V_{ks}|^2 \int_0^{\Delta t} (\Delta t - t') e^{i(\omega_s - \omega_k)t'} \, dt' \ldots + \sum_k |V_{ks}|^2 \int_{(j-1)\Delta t}^{j\Delta t} (j\Delta t - t') e^{i(\omega_s - \omega_k)(t' - [j-1]\Delta t)} \, dt' \right. \tag{13}$$

$$\left. \ldots\ldots\ldots + \sum_k |V_{ks}|^2 \int_{(2n-1)\Delta t}^{2n\Delta t} (2n\Delta t - t') e^{i(\omega_s - \omega_k)(t' - [2n-1]\Delta t)} \, dt' \right)$$

and changing the limits of integration, in Eq. (13), we obtain:

$$\text{Terms from } I = 2n\Delta t \sum_k |V_{ks}|^2 \int_0^{\Delta t} (1 - \frac{t'}{\Delta t}) e^{i(\omega_s - \omega_k)t'} \, dt' \, . \tag{14}$$

Similarly, contributions from terms *II* and *III* are obtain, as follows:

$$\text{Terms from } II \text{ and } III = \sum_k \sum_{j=1}^{2n-1} (-1)^j \int_{j\Delta t}^{\{j+1\}\Delta t} V_{sk} e^{i(\omega_s - \omega_k)t'} \beta_k(j\Delta t) \, dt' \tag{15}$$

where the summation over $j$ starts at $j=1$ because $\beta_k(0) = 0$. Substituting Eqs. (14) and (15) into Eq. (12), we obtain:

$$\alpha(2n\Delta t) = 1 - 2n\Delta t \int_0^{\Delta t} (1 - \frac{t'}{\Delta t}) K(t') \, dt' - i \sum_k \sum_{j=1}^{2n-1} (-1)^j \int_{j\Delta t}^{\{j+1\}\Delta t} V_{sk} e^{i(\omega_s - \omega_k)t'} \beta_k(j\Delta t) \, dt'$$

$$= 1 - 2n\Delta t \int_0^{\Delta t} (1 - \frac{t'}{\Delta t}) K(t') \, dt' - i \sum_k \sum_{j=1}^{2n-1} (-1)^j V_{sk} \left( \frac{e^{i(\omega_s - \omega_k)\Delta t} - 1}{i(\omega_s - \omega_k)} \right) \beta_k(j\Delta t) \tag{16}$$

Appendix B shows that one can substitute,

$$\beta_k(j\Delta t) = \frac{V_{ks}}{(\omega_k - \omega_s)} \frac{e^{i(\omega_k - \omega_s)\Delta t} - 1}{e^{i(\omega_k - \omega_s)\Delta t} + 1} \left\{ (-1)^j e^{i(\omega_k - \omega_s)j\Delta t} - 1 \right\} \tag{17}$$

into Eq. (16) to obtain

$$\alpha_s(2n\Delta t) = 1 - 2n\Delta t \int_0^{\Delta t} (1 - \frac{t'}{\Delta t}) K(t') \, dt'$$

$$- \sum_k e^{i(\omega_s - \omega_k)\Delta t} \frac{|V_{ks}|^2}{(\omega_k - \omega_s)^2} \frac{\left(e^{i(\omega_k - \omega_s)\Delta t} - 1\right)^2}{e^{i(\omega_k - \omega_s)\Delta t} + 1} \sum_{j=1}^{2n-1} \left\{ 1 - (-1)^j e^{i(\omega_s - \omega_k)j\Delta t} \right\} \tag{18}$$

and substituting $\sum_{j=1}^{2n-1} R^j = \frac{R(1-R^{2n-1})}{(1-R)}$ with $R \equiv -e^{i(\omega_s - \omega_k)\Delta t}$ into Eq. (18), gives

$$\begin{aligned}
\alpha_s(2n\Delta t) =\ & 1 - 2n\Delta t \int_0^{\Delta t} (1 - \frac{t'}{\Delta t}) \mathrm{K}(t')\, dt' \\
& - \sum_k e^{i(\omega_s - \omega_k)\Delta t} \frac{|V_{ks}|^2}{(\omega_k - \omega_s)^2} \frac{\left(e^{i(\omega_k - \omega_s)\Delta t} - 1\right)^2}{e^{i(\omega_k - \omega_s)\Delta t} + 1} \left( 2n - 1 + \frac{e^{i(\omega_s - \omega_k)(2n)\Delta t} + e^{i(\omega_s - \omega_k)\Delta t}}{e^{i(\omega_s - \omega_k)\Delta t} + 1} \right) \\
=\ & 1 - 2n\Delta t \int_0^{\Delta t} (1 - \frac{t'}{\Delta t}) \mathrm{K}(t')\, dt' \\
& - \sum_k e^{i(\omega_s - \omega_k)\Delta t} \frac{|V_{ks}|^2}{(\omega_k - \omega_s)^2} \frac{\left(e^{i(\omega_k - \omega_s)\Delta t} - 1\right)^2}{e^{i(\omega_k - \omega_s)\Delta t} + 1} \left( 2n + \frac{e^{i(\omega_s - \omega_k)(2n)\Delta t} - 1}{e^{i(\omega_s - \omega_k)\Delta t} \left( e^{i(\omega_k - \omega_s)\Delta t} + 1 \right)} \right) \\
=\ & 1 - 2n\Delta t \int_0^{\Delta t} (1 - \frac{t'}{\Delta t}) \mathrm{K}(t')\, dt' \ - \sum_k \underbrace{\frac{|V_{sk}|^2}{(\omega_k - \omega_s)^2} \left( \frac{e^{i(\omega_k - \omega_s)\Delta t} - 1}{e^{i(\omega_k - \omega_s)\Delta t} + 1} \right)^2 \left( e^{-i(\omega_k - \omega_s)(2n)\Delta t} - 1 \right)}_{F_k^1} \\
& - 2n \sum_k \underbrace{\frac{|V_{sk}|^2}{(\omega_k - \omega_s)^2} e^{-i(\omega_k - \omega_s)\Delta t} \frac{\left(e^{i(\omega_k - \omega_s)\Delta t} - 1\right)^2}{(e^{i(\omega_k - \omega_s)\Delta t} + 1)}}_{F_k^2} \\
=\ & 1 - 2n\Delta t \int_0^{\Delta t} (1 - \frac{t'}{\Delta t}) \mathrm{K}(t')\, dt' \ - \sum_k F_k^1 - \sum_k F_k^2
\end{aligned} \tag{19}$$

Equation (19) is an important result since it provides an explicit description of the state amplitude $\alpha_s(2n\Delta t)$ as a function of the time interval $\Delta t$ between phase-kick pulses, yielding fundamental insight on the origin of interference phenomena due to the various terms. In addition, Eq. (19) allows for calculations of the survival probability of state $|s\rangle$:

$$|\alpha_s(2n\Delta t)|^2 = 1 - \underbrace{(2n\Delta t) 2\mathrm{Re} \int_0^{\Delta t} (1 - \frac{t'}{\Delta t}) \mathrm{K}(t')\, dt'}_{A} - \underbrace{2\mathrm{Re} \sum_k F_k^1}_{B} - \underbrace{2\mathrm{Re} \sum_k F_k^2}_{C} + \cdots \tag{20}$$

with

$$2\mathrm{Re} \int_0^{\Delta t} \left(1 - \frac{t'}{\Delta t}\right) \mathrm{K}(t')\, dt' = 2\mathrm{Re} \sum_k |V_{ks}|^2 \int_0^{\Delta t} \left(1 - \frac{t'}{\Delta t}\right) e^{i(\omega_s - \omega_k)t'}\, dt' = \Delta t \sum_k |V_{ks}|^2 \frac{\sin^2\left((\omega_s - \omega_k)\frac{\Delta t}{2}\right)}{\left((\omega_s - \omega_k)\frac{\Delta t}{2}\right)^2}, \tag{21}$$

$$2\operatorname{Re}\sum_k F_k^1 = \sum_k \frac{|V_{ks}|^2}{\left(\frac{\omega_s - \omega_k}{2}\right)^2} \tan^2\left((\omega_s - \omega_k)\frac{\Delta t}{2}\right) \sin^2\left((\omega_s - \omega_k)\frac{2n\Delta t}{2}\right), \quad (22)$$

and

$$2\operatorname{Re}\sum_k F_k^2 = -2n\sum_k \frac{|V_{ks}|^2}{\left(\frac{\omega_s - \omega_k}{2}\right)^2} \sin^2(\omega_s - \omega_k)\frac{\Delta t}{2}. \quad (23)$$

Note that terms A and C in Eq. (20) cancel each other and term B determines the time-dependent survival probability, as follows:

$$|\alpha_s(2n\Delta t)|^2 = 1 - 2\operatorname{Re}\sum_k F_k^1 = 1 - \sum_k \frac{|V_{ks}|^2}{\left(\frac{\omega_s - \omega_k}{2}\right)^2} \tan^2\left((\omega_s - \omega_k)\frac{\Delta t}{2}\right) \sin^2\left((\omega_s - \omega_k)\frac{2n\Delta t}{2}\right). \quad (24)$$

Equation (24) gives the survival probability $P_s(t) = |\alpha_s(t)|^2$, as a function of the time-interval $\Delta t$ between pulses. Due to the functional dependence, $\tan^2((\omega_s - \omega_k)\Delta t/2)$, decay is suppressed when the time interval between pulses is sufficiently short $\Delta t \to 0$ (with $t = 2n\Delta t$), and is accelerated relative to spontaneous decay, when $\tan^2((\omega_s - \omega_k)\Delta t/2) > 1$ (see, Eq. (9) as compared to Eq. (24)). Maximum acceleration is achieved when $\Delta t = \pi/(\omega_s - \omega_k)$. Equation (24) is consistent with previous work,[23,24] including the study of decay into a continuum,[27,29,44] and the description of the decay of coherence in a system of spin ½ qubits, in contact with a bosonic bath, when periodically pulsed by dynamical decoupling sequences.[29] However, its derivation is novel since, contrary to earlier studies, it is derived from Eq. (19) providing an explicit description of the evolution of the expansion coefficient $\alpha_s(t)$ as a function of the time-interval $\Delta t$ between pulses.

## IV. STOCHASTIC PULSING

This section analyzes stochastic sequences of $2\pi$ pulses, using the perturbational treatment introduced in Sec. III. Rather that pulsing the system deterministically, as in Sec. III, stochastic sequences pulse the system at time intervals $\Delta t$ but only with 50% probability.

To obtain the survival probability $P_s(t) = |\alpha_s(t)|^2$, at time $t = 2n\Delta t$, we analyze first the state of the system at time $t = 2\Delta t$, obtained by propagating the expansion coefficients for states $|s\rangle$ and $|k\rangle$, as follows:

$$\beta_k(\Delta t) = \beta_k(0) - iV_{ks}\left(\frac{e^{i(\omega_k - \omega_s)\Delta t} - 1}{i(\omega_k - \omega_s)}\right)\alpha_s(0)$$

$$\alpha_s(\Delta t) = \alpha_s(0)\left(1 - \int_0^{\Delta t}(\Delta t - t')K(t')\,dt'\right) - i\sum_k V_{sk}\int_0^{\Delta t} e^{i(\omega_s - \omega_k)t}\beta_k(0)\,dt$$

$$\alpha'_s(\Delta t) = \xi_1 \alpha_s(\Delta t) \qquad (25)$$

$$\beta_k(2\Delta t) = \beta_k(\Delta t) - iV_{ks}\left(\frac{e^{i(\omega_k - \omega_s)2\Delta t} - e^{i(\omega_k - \omega_s)\Delta t}}{i(\omega_k - \omega_s)}\right)\alpha'_s(\Delta t)$$

$$\alpha'_s(2\Delta t) = \alpha'_s(\Delta t)\left(1 - \int_{\Delta t}^{2\Delta t}(2\Delta t - t')K(t' - \Delta t)\,dt'\right) - i\sum_k V_{sk}\int_{\Delta t}^{2\Delta t} e^{i(\omega_s - \omega_k)t}\beta_k(\Delta t)\,dt$$

$$\alpha_s(2\Delta t) = \xi_2 \alpha'_s(2\Delta t)$$

where $\xi_j$ are stochastic variables that take on values of $\pm 1$, with equal probability, and correspond to the system being perturbed (*i.e.*, $\xi_j = -1$) by a $2\pi$ pulse (*i.e.*, $\hat{Q} = 1 - |s\rangle\langle s|$) at time $t_j = j\Delta t$, or not (*i.e.*, $\xi_j = 1$). A comparative analysis of the resulting stochastic sequence to a dynamical decoupling scheme based on random variables random variables are $\chi_n = \{(-1)^n, n \in \mathbb{N}\}$, previously considered by Santos and Viola for manipulating coherence in spin ½ qubits,[41,42] is discussed in Appendix C.

The expansion coefficients for the continuum states are obtained, as follows:

$$\beta_k(l\Delta t) = -iV_{ks}\left(\frac{e^{i(\omega_k-\omega_s)\Delta t}-1}{i(\omega_k-\omega_s)}\right)\left(1+\sum_{j=1}^{l-1}\xi_j e^{i(\omega_k-\omega_s)j\Delta t}\right) \quad (26)$$

and the time-evolution of the initially populated state $|s\rangle$ is,

$$\alpha_s(2n\Delta t) = \underbrace{\prod_{j=1}^{2n}\xi_j\left(1-2n\Delta t\int_0^{\Delta t}(1-\frac{t'}{\Delta t})\mathrm{K}(t')\,dt'\right)}_{G}$$

$$\underbrace{-\sum_k\sum_{l=1}^{2n-1}\prod_{j=l}^{2n}\xi_j\left(e^{-i(\omega_k-\omega_s)[l+1]\Delta t}-e^{-i(\omega_k-\omega_s)l\Delta t}\right)\frac{|V_{ks}|^2}{(\omega_k-\omega_s)^2}\left(e^{i(\omega_k-\omega_s)\Delta t}-1\right)\left(1+\sum_{i=1}^{l-1}\xi_i e^{i(\omega_k-\omega_s)i\Delta t}\right)}_{F} \quad (27)$$

Note that in the limit when $\xi_j \to 1$ (*i.e.*, pulses with 0% efficiency), Eq. (27) yields,

$$|\alpha(2n\Delta t)|^2 = 1 - \sum_k \frac{|V_{ks}|^2}{\left(\frac{\omega_s-\omega_k}{2}\right)^2}\sin^2\left((\omega_s-\omega_k)\frac{2n\Delta t}{2}\right), \quad (28)$$

that is the expression for spontaneous decay, introduced by Eq. (0.9).[34] More generally, the survival probability of the system evolving under the effect of pulses with $\xi_j \neq 1$ is:

$$|\alpha(2n\Delta t)|^2 = |G|^2 - 2\mathrm{Re}(F^*G) + |F|^2$$

$$|G|^2 = 1 - 2\mathrm{Re}\left(2n\Delta t\int_0^{\Delta t}(1-\frac{t'}{\Delta t})\mathrm{K}(t')dt'\right)$$

$$F^*G = \prod_{j=1}^{2n}\xi_j\sum_k\sum_{l=1}^{2n-1}\prod_{a=l}^{2n}\xi_a\left(\frac{e^{-i(\omega_k-\omega_s)[l+1]\Delta t}-e^{-i(\omega_k-\omega_s)l\Delta t}}{i(\omega_s-\omega_k)}\right)\times \quad (29)$$

$$\frac{|V_{ks}|^2}{(\omega_k-\omega_s)^2}\left(e^{i(\omega_k-\omega_s)\Delta t}-1\right)(1+\sum_{b=1}^{l-1}\xi_b e^{i(\omega_k-\omega_s)b\Delta t})$$

where $|F|^2$ is neglected since it involves terms of $O(|V_{ks}|^4)$. Equation (29) shows that coherent-control can be achieved with stochastic sequences of phase-kick pulses. Note that the population decay is suppressed (*i.e.*, $|\alpha(2n\Delta t)|^2 \to 1$) when $\Delta t \to 0$. In addition, decay can be accelerated, relative to the spontaneous behavior described by Eq. (9), for larger values of $\Delta t$.

To analyze the effect of averaging over all possible stochastic sequences, we consider independent random variables with $\langle \xi_j \rangle = 0$ and

$$\langle \xi_1 \xi_2 \ldots \xi_n \rangle = \langle \xi_1 \rangle \langle \xi_2 \rangle \langle \xi_3 \rangle \ldots \langle \xi_n \rangle = 0. \tag{30}$$

Therefore, $\langle F^*G \rangle = 0$ and the average short-time decay rate at $t = 2n\Delta t$ is:

$$\langle |\alpha(2n\Delta t)|^2 \rangle = |G|^2$$

$$|G|^2 = 1 - 2n\Delta t \left\{ 2\operatorname{Re}\left( \int_0^{\Delta t} (1 - \frac{t'}{\Delta t}) K(t') dt' \right) \right\} = 1 - \gamma_{avg} 2n\Delta t \tag{31}$$

$$\gamma_{avg} = 2\operatorname{Re}\left( \int_0^{\Delta t} (1 - \frac{t'}{\Delta t}) K(t') dt' \right) = \Delta t \sum_k |V_{ks}|^2 \frac{\sin^2\left((\omega_k - \omega_s)\frac{\Delta t}{2}\right)}{\left((\omega_k - \omega_s)\frac{\Delta t}{2}\right)^2}$$

Interestingly, $\gamma_{avg}$ is exactly the decay rate derived by Kofman and Kurizki in the context of QZE,[34-36] where contrary to unitary phase-kick pulses, the pulses *collapse* the coherent evolution, as due to a measurement, by projecting the time-evolved state into a state (*e.g.*, $|s\rangle$). For comparison, Sec. V derives the QZE and AZE dynamics by using the perturbational treatment implemented in this section in conjunction with pulses that collapse the coherent evolution into state $|s\rangle$. The observed correspondence in the decay rates suggests that the dynamical effect of repetitive measurements is equivalent to the average effect of stochastic phase-kick pulses when averaged over all possible realizations.

**V. QUANTUM ZENO AND QUANTUM ANTI-ZENO EFFECTS**

Quantum Zeno and Anti-Zeno effects (QZE and AZE) occur if the coherent evolution of the system state is interrupted by a sequence of time–periodic measurements.[32] If the process is interrupted sufficiently frequently one observes a complete freezing of decay dynamics (Zeno effect),[31] and with longer time intervals between pulses acceleration of decay (Anti-Zeno effect)[34] is observed. In their landmark work on the topic, Kofman and Kurizki elucidated the mechanism via which both these effects set in, hinting at the relation between the density of states of the continuum and the time interval between measurements. We refer the reader to the original work of Kofman and Kurizki[34] for the relevant details of the processes. In this section, we highlight the quantitative similarities of the process with the pulsing coherent control schemes, described in Secs. III and IV.

We consider the Hamiltonian, introduced by Eq. (1), with $\hat{U}$ denoting the short-time evolution as described by Eq. (7) and Eq. (8). $\hat{P} = |s\rangle\langle s|$ represents the measurement process, collapsing the system onto state $|s\rangle$ at time $\Delta t$ and yielding a state with

$$\alpha_s(\Delta t) = \langle s | \hat{P}\hat{U} | \psi \rangle \tag{32}$$

and devoid of any population in states $|k\rangle$. Now, if the time evolution proceeds in sufficiently small time steps of order $\Delta t$ and $\beta_k(0) = 0$, then the population of states $|k\rangle$ will remain negligible for later times. Using Equation (7) for computing the survival probability in state $|s\rangle$, we obtain:

$$|\alpha_s(\Delta t)|^2 = |\alpha_s(0)|^2 \left( 1 - 2\,\text{Re}\left\{ \sum_k |V_{ks}|^2 \int_0^{\Delta t} (\Delta t - t') e^{i(\omega_s - \omega_k)t'} dt' \right\} \right) \tag{33}$$

Repeating the measurement $2n$ times, we obtain the survival probability at time $t = 2n\Delta t$:

$$|\alpha_s(2n\Delta t)|^2 = |\alpha_s([2n-1]\Delta t)|^2 \left(1 - 2\operatorname{Re}\left\{\sum_k |V_{ks}|^2 \int_0^{\Delta t} (\Delta t - t')e^{i(\omega_s - \omega_k)t'} dt'\right\}\right) \quad (34)$$

Substituting Eq. (33) into Eq. (34) recursively, we obtain the survival probability at $t = 2n\Delta t$,

$$\begin{aligned}|\alpha_s(2n\Delta t)|^2 &= |\alpha_s(0)|^2 \left(1 - 2\operatorname{Re}\left\{\sum_k |V_{ks}|^2 \int_0^{\Delta t} (\Delta t - t')e^{i(\omega_s - \omega_k)t'} dt'\right\}\right)^{2n} \\ &\approx |\alpha_s(0)|^2 \left(1 - 2\operatorname{Re}\left\{\sum_k |V_{ks}|^2 2n\Delta t \int_0^{\Delta t} (1 - \frac{t'}{\Delta t})e^{i(\omega_s - \omega_k)t'} dt'\right\}\right)\end{aligned} \quad (35)$$

where we have neglected terms of $o(|V_{ks}|^3)$ and higher, as appropriate in the weak coupling limit. After the final integration, the above expression takes the form:

$$\begin{aligned}|\alpha_s(2n\Delta t)|^2 &= |\alpha_s(0)|^2 \left(1 - 2\operatorname{Re}\left\{\sum_k |V_{ks}|^2 \int_0^{\Delta t} (\Delta t - t')e^{i(\omega_s - \omega_k)t'} dt'\right\}\right)^{2n} \\ &\approx |\alpha_s(0)|^2 \left(1 - 2n\Delta t \sum_k |V_{ks}|^2 \Delta t \underbrace{\frac{\sin^2\left(\{\omega_k - \omega_s\}\frac{\Delta t}{2}\right)}{\left(\{\omega_k - \omega_s\}\frac{\Delta t}{2}\right)^2}}_{\gamma_{ZENO}}\right)\end{aligned} \quad (36)$$

The rate $\gamma_{ZENO}$ is identical to term A in the expression of the survival probability for the system under the pulsed coherent evolution (see Eq. (20)). Such a term A, therefore, leads to the effective emergence of QZE and AZE when terms B and C cancel.

**VI. CONCLUSIONS**

In this paper we have shown that quantum tunneling can be suppressed, or accelerated, by using deterministic, or stochastic, sequences of unitary pulses that affect the underlying interference phenomena responsible for quantum dynamics, without inducing decoherence, or collapsing the coherent evolution of the system. A rigorous theoretical analysis, based on perturbation theory to first order in the control pulse fields, showed that sufficiently frequent perturbation pulses suppress quantum tunneling while trains of pulses separated by finite time intervals accelerate tunneling relative to spontaneous decay. The reported expressions also provided understanding on the role of randomization and the emergence of dynamics analogous to the evolution due to QZE or AZE, generated by stochastic sequences of unitary pulses when averaged over all possible realizations. The comparison to DD protocols and control schemes based on pulses that collapse the coherent evolution reveals a subtle interplay between coherent and incoherent phenomena that can be exploited by averaging stochastic sequences of unitary pulses when averaged over all possible realizations. However, we emphasize that the resulting coherent control induced by stochastic or deterministic sequences of unitary pulses is due to interference phenomena associated with quantum dynamics in between pulses as described by the unperturbed Hamiltonian.

 Our theoretical procedure showed how to analyze coherent control techniques based on sequences of unitary pulses, QZE, AZE and DD techniques on an equal mathematical footing. The calculations essentially unify the treatments due to Kofman, Kurizki[23,24] and Agarwal *et. al.* [23,24,29,44] and in  the process go beyond their treatments to reveal the inherent intricacies of the dynamics and shows that the decay pattern for deterministic decoupling is not restricted to a particular system (e.g., a system of spin ½ qubits) but in essence is of universal nature.[27] This assertion is supported by the analysis of a common system, tunneling to a continuum, as affected by the various control techniques.

Our theoretical analysis has shown that QZE terms such as terms A and G in Eq. (19) and Eq. (27), respectively, are included even in the expressions of state amplitudes affected by coherent control sequences based on unitary pulses. However, only the judicious tailoring of pulses that affects the interference of the system state with the continuum states leads to the manifestation of QZE. The emergence of such behavior upon random pulsing is due to the stochastic phase that washes out the coherent interference effects and brings forward the otherwise suppressed incoherent effects.

Considering the simplicity of sequences based on phase-kick pulses, the similarity to pulsed NMR techniques, and the fact that other pulse sequences have already been demonstrated to achieve control in condensed material systems, we anticipate that the control techniques analyzed in this paper should raise significant experimental interest.

## ACKNOWLEDGEMENTS

The authors are grateful to Prof. Shaul Mukamel for teaching them the beauty of quantum mechanics, as applied to the description pulse-matter interactions. We thank Prof. Lea F. Santos (Yeshiva University) for helpful comments on a preliminary version of this manuscript. V.S.B. acknowledges supercomputer time from NERSC and support from NSF grants CHE-0911520 and ECCS-040419. Preliminary work on quantum dynamics for coherent control has been funded by the Division of Chemical Sciences, Geosciences, and Biosciences, Office of Basic Energy Sciences of the U.S., Department of Energy (DE-FG02-07ER15909).

**APPENDIX A**

This section derives the short-time approximation, introduced by Eq. (7). For a sufficiently short time-interval ($t$-$t_b$), we assume $\alpha_s(t') \approx \alpha_s(t_b)$ in Eq. (6):

$$\begin{aligned}
\dot{\alpha}_s(t) &\approx -\alpha_s(t_b) \int_{t_b}^{t} \sum_k |V_{ks}|^2 e^{i(\omega_s - \omega_k)(t-t')} \, dt' - i \sum_k V_{sk} e^{i(\omega_s - \omega_k)t} \beta_k(t_b) \\
&= \alpha_s(t_b) \sum_k |V_{ks}|^2 \frac{1 - e^{i(\omega_s - \omega_k)(t-t_b)}}{i(\omega_s - \omega_k)} - i \sum_k V_{sk} e^{i(\omega_s - \omega_k)t} \beta_k(t_b)
\end{aligned} \tag{A.1}$$

Integrating Eq. (A.1) by parts, we obtain:

$$\begin{aligned}
\alpha_s(t) - \alpha_s(t_b) &= \alpha_s(t_b) \sum_k |V_{ks}|^2 t' \frac{1 - e^{i(\omega_s - \omega_k)(t'-t_b)}}{i(\omega_s - \omega_k)} \bigg|_{t_b}^{t} + \int_{t_b}^{t} t' e^{i(\omega_s - \omega_k)(t'-t_b)} dt' - i \sum_k V_{sk} \int_{t_b}^{t} dt' e^{i(\omega_s - \omega_k)t'} \beta_k(t_b) \\
&= -\alpha_s(t_b) \sum_k |V_{ks}|^2 t \int_{t_b}^{t} e^{i(\omega_s - \omega_k)(t-t_b)} dt' + \int_{t_b}^{t} t' e^{i(\omega_s - \omega_k)(t'-t_b)} dt' - i \sum_k V_{sk} \int_{t_b}^{t} dt' e^{i(\omega_s - \omega_k)t'} \beta_k(t_b) \\
&= -\alpha_s(t_b) \sum_k |V_{ks}|^2 \int_{t_b}^{t} (t-t') e^{i(\omega_s - \omega_k)(t'-t_b)} dt' - i \sum_k V_{sk} \int_{t_b}^{t} dt' e^{i(\omega_s - \omega_k)t'} \beta_k(t_b)
\end{aligned} \tag{A.2}$$

**APPENDIX B**

Using, Eq. (5) and the scheme defined in Eq. (11), the evolution of the continuum states, in steps $\Delta t$ is obtained as follows:

$$\beta_k(\Delta t) = \beta_k(0) - iV_{ks}\left(\frac{e^{i(\omega_k-\omega_s)\Delta t}-1}{i(\omega_k-\omega_s)}\right)\alpha_s(0)$$

$$\beta_k(2\Delta t) = \beta_k(0) - iV_{ks}\left(\frac{e^{i(\omega_k-\omega_s)\Delta t}-1}{i(\omega_k-\omega_s)}\right)\alpha_s(0) - iV_{ks}\left(\frac{e^{i(\omega_k-\omega_s)2\Delta t}-e^{i(\omega_k-\omega_s)\Delta t}}{i(\omega_k-\omega_s)}\right)\alpha'_s(\Delta t)$$

$$\beta_k(2n\Delta t) = \beta_k(0) - iV_{ks}\left(\frac{e^{i(\omega_k-\omega_s)\Delta t}-1}{i(\omega_k-\omega_s)}\right)\alpha_s(0) - iV_{ks}\left(\frac{e^{i(\omega_k-\omega_s)2\Delta t}-e^{i(\omega_k-\omega_s)\Delta t}}{i(\omega_k-\omega_s)}\right)\alpha'_s(\Delta t) -$$

$$-iV_{ks}\left(\frac{e^{i(\omega_k-\omega_s)3\Delta t}-e^{i(\omega_k-\omega_s)2\Delta t}}{i(\omega_k-\omega_s)}\right)\alpha_s(2\Delta t) + iV_{ks}\left(\frac{e^{i(\omega_k-\omega_s)4\Delta t}-e^{i(\omega_k-\omega_s)3\Delta t}}{i(\omega_k-\omega_s)}\right)\alpha'_s(3\Delta t) \ldots$$

$$\ldots\ldots iV_{ks}\left(\frac{e^{i(\omega_k-\omega_s)(2n-1)\Delta t}-e^{i(\omega_k-\omega_s)(2n-2)\Delta t}}{i(\omega_k-\omega_s)}\right)\alpha_s([2n-2]\Delta t)$$

(B.1)

where $\alpha'_s(l\Delta t') = -\alpha_s(l\Delta t)$ accounts for the phase flip due to the action of a $2\pi$ pulse. To obtain an expression of $\beta_k(2n\Delta t)$ of $O\left(|V_{ks}|^2\right)$ we keep only the zero order term in the expansions of $\alpha_s(l\Delta t)$ in powers of $V_{ks}$ and we obtain the compact expressions for the continuum state amplitudes, as follows:

$$\beta_k(\Delta t) = \beta_k(0) - iV_{ks}\left(\frac{e^{i(\omega_k - \omega_s)\Delta t} - 1}{i(\omega_k - \omega_s)}\right)\alpha_s(0)$$

$$\beta_k(2\Delta t) = \beta_k(0) - iV_{ks}\left(\frac{e^{i(\omega_k - \omega_s)\Delta t} - 1}{i(\omega_k - \omega_s)}\right)\alpha_s(0) \; - \; iV_{ks}\left(\frac{e^{i(\omega_k - \omega_s)2\Delta t} - e^{i(\omega_k - \omega_s)\Delta t}}{i(\omega_k - \omega_s)}\right)\alpha_s'(0)$$

$$\beta_k([2n-1]\Delta t) = \beta_k(0) - iV_{ks}\left(\frac{e^{i(\omega_k - \omega_s)\Delta t} - 1}{i(\omega_k - \omega_s)}\right)\alpha_s(0) \; - \; iV_{ks}\left(\frac{e^{i(\omega_k - \omega_s)2\Delta t} - e^{i(\omega_k - \omega_s)\Delta t}}{i(\omega_k - \omega_s)}\right)\alpha_s'(0) -$$

$$-iV_{ks}\left(\frac{e^{i(\omega_k - \omega_s)3\Delta t} - e^{i(\omega_k - \omega_s)2\Delta t}}{i(\omega_k - \omega_s)}\right)\alpha_s(0) + iV_{ks}\left(\frac{e^{i(\omega_k - \omega_s)4\Delta t} - e^{i(\omega_k - \omega_s)3\Delta t}}{i(\omega_k - \omega_s)}\right)\alpha_s'(0) \ldots$$

$$\ldots\ldots iV_{ks}\left(\frac{e^{i(\omega_k - \omega_s)(2n-1)\Delta t} - e^{i(\omega_k - \omega_s)(2n-2)\Delta t}}{i(\omega_k - \omega_s)}\right)\alpha_s(0)$$

$$\beta_k(l\Delta t) = \frac{V_{ks}}{(\omega_k - \omega_s)}\sum_{j=1}^{l}(-1)^j\left(e^{i*j*(\omega_k - \omega_s)*\Delta t} - e^{i*(j-1)*(\omega_k - \omega_s)*\Delta t}\right)$$

$$= \frac{V_{ks}}{(\omega_k - \omega_s)}\frac{(e^{i(\omega_k - \omega_s)\Delta t} - 1)((-1)^l e^{i(\omega_k - \omega_s)l\Delta t} - 1)}{e^{i(\omega_k - \omega_s)\Delta t} + 1} \qquad (\because \beta_k(0) = 0) \tag{B.2}$$

Note that the continuum-state amplitude at any particular time step, accounts for all the continuum state amplitudes at prior time steps. Moreover, contributions from even and odd time steps occur with alternating signs. This is a direct consequence of the phase flip of the system state as a result of the pulsing which affects the above evolution equations in the form of $\alpha_s(0)$. An interesting analogy emerges if one interprets the sign change as a time reversal of continuum dynamics under successive pulse applications.[45] In the context of NMR, this amounts to a spin echoes [29] initiated with the purpose of negating the continuum induced decoherence in spin-spin correlations. In, the event of no pulses, the above expression becomes a telescoping sum, which eventually leads to the spontaneous decay behavior.

**APPENDIX C**

This section compares the coherent control scenario based on random variables $\xi_n = \pm 1$, introduced in Sec. IV, to the dynamical decoupling scheme based on random variables $\chi_n = \{(-1)^n, n \in \mathbb{N}\}$ considered by Santos and Viola for manipulating coherence in spin ½ qubits.[29] We modify the scheme defined in Eq. (25), as follows:

$$\beta_k(\Delta t) = \beta_k(0) - iV_{ks}\lambda_0 \left( \frac{e^{i(\omega_k - \omega_s)\Delta t} - 1}{i(\omega_k - \omega_s)} \right) \alpha_s(0)$$

$$\alpha_s(\Delta t) = \alpha_s(0)\left(1 - \int_0^{\Delta t}(\Delta t - t')\mathrm{K}(t')\,dt'\right) - i\sum_k V_{sk}\int_0^{\Delta t} e^{i(\omega_s - \omega_k)t}\lambda_0 \beta_k(0)\,dt$$

$$\beta_k(2\Delta t) = \beta_k(\Delta t) - iV_{ks}\lambda_1 \left( \frac{e^{i(\omega_k - \omega_s)2\Delta t} - e^{i(\omega_k - \omega_s)\Delta t}}{i(\omega_k - \omega_s)} \right) \alpha_s(\Delta t) \quad \text{(C.1)}$$

$$\alpha_s(2\Delta t) = \alpha_s(\Delta t)\left(1 - \int_{\Delta t}^{2\Delta t}(2\Delta t - t')\mathrm{K}(t' - \Delta t)\,dt'\right) - i\sum_k V_{sk}\int_{\Delta t}^{2\Delta t} e^{i(\omega_s - \omega_k)t}\lambda_1 \beta_k(\Delta t)\,dt$$

where $K(t) = \sum_k |V_{ks}|^2 e^{i(\omega_s - \omega_k)t}$ and $\lambda_j = (-1)^j$ for a deterministic pulsing scheme. If we collect the expressions from Eq. (C.1) we obtain

$$\alpha_s(2\Delta t) = \left\{\alpha_s(0)\left(1 - \int_0^{\Delta t}(\Delta t - t')\mathrm{K}(t')\,dt'\right) - i\sum_k \int_0^{\Delta t} V_{sk} e^{i(\omega_s - \omega_k)t'}\lambda_0 \beta_k(0)\,dt'\right\} \times$$

$$\left\{\left(1 - \int_{\Delta t}^{2\Delta t}(2\Delta t - t')\mathrm{K}(t' - \Delta t)\,dt'\right)\right\} - i\sum_k \int_{\Delta t}^{2\Delta t} V_{sk} e^{i(\omega_s - \omega_k)t'}\lambda_1 \beta_k(\Delta t)\,dt'$$

$$= 1 - \int_0^{\Delta t}(\Delta t - t')\mathrm{K}(t')\,dt' - \int_{\Delta t}^{2\Delta t}(2\Delta t - t')\mathrm{K}(t' - \Delta t)\,dt'$$

$$- i\sum_k \int_0^{\Delta t} V_{sk} e^{i(\omega_s - \omega_k)t'}\lambda_0 \beta_k(0)\,dt' - i\sum_k \int_{\Delta t}^{2\Delta t} V_{sk} e^{i(\omega_s - \omega_k)t'}\lambda_1 \beta_k(\Delta t)\,dt' \quad \text{(C.2)}$$

$$= 1 - 2\int_0^{\Delta t}(\Delta t - t')\mathrm{K}(t')\,dt'$$

$$- i\sum_k \int_0^{\Delta t} V_{sk} e^{i(\omega_s - \omega_k)t'}\lambda_0 \beta_k(0)\,dt' - i\sum_k \int_{\Delta t}^{2\Delta t} V_{sk} e^{i(\omega_s - \omega_k)t'}\lambda_1 \beta_k(\Delta t)\,dt'$$

and

$$\beta_k(l\Delta t) = -iV_{ks}\left(\frac{e^{i(\omega_k-\omega_s)\Delta t}-1}{i(\omega_k-\omega_s)}\right)\left(\sum_{j=0}^{l-1}\lambda_j e^{i(\omega_k-\omega_s)j\Delta t}\right) \quad (C.3)$$

It can be verified that using the above definition for continuum states and substituting it back into Eq. (C.1)) one obtains exactly the expression derived in Eq. (26). We see from Eq. (C.3) that the continuum state amplitude is a combination of terms that alternate in sign as in Eq. (B.2) of Appendix B. Going back to our analogy of pulse applications and spin echoes (see Appendix B), the application of periodic $2\pi$ pulses is equivalent to the initiation of successive $\pi$ phase shifts in the continuum state amplitude. This is accomplished in this case by allowing the variables to be $\lambda_j = (-1)^j$. Using these definitions, the expression for the survival amplitude becomes:

$$\alpha(2n\Delta t) = \left(1 - 2n\Delta t\int_0^{\Delta t}(1-\frac{t'}{\Delta t})\mathrm{K}(t')\,dt'\right) - i\sum_k\sum_{l=1}^{2n-1}\int_{l\Delta t}^{\{l+1\}\Delta t}V_{sk}e^{i(\omega_s-\omega_k)t'}\lambda_l\beta_k(l\Delta t)dt' \quad (C.4)$$

Consequently, the resulting survival amplitude is

$$|\alpha(2n\Delta t)|^2 = |G|^2 + 2\mathrm{Re}(F^*G) + |F|^2$$

$$|G|^2 = 1 - 2\mathrm{Re}\left(2n\Delta t\int_0^{\Delta t}(1-\frac{t'}{\Delta t})K(t')dt'\right) = 1 - 2n\Delta t\sum_k\Delta t\frac{\sin^2(\omega_s-\omega_k)\frac{\Delta t}{2}}{\left[\frac{(\omega_s-\omega_k)\Delta t}{2}\right]^2}|V_{ks}|^2$$

$$2\mathrm{Re}(F^*G) = 2\mathrm{Re}\left(\sum_k\sum_{l=1}^{2n-1}\sum_{m=0}^{l-1}\lambda_l\lambda_m\frac{|V_{ks}|^2}{(\omega_k-\omega_s)^2}\left(e^{i(\omega_k-\omega_s)\Delta t}-1\right)^2 e^{-i(\omega_k-\omega_s)(l-m+1)\Delta t}\right)$$

$$= -2\mathrm{Re}\left(\sum_k\sum_{l=1}^{2n-1}\sum_{m=0}^{l-1}\lambda_l\lambda_m\frac{|V_{ks}|^2}{(\omega_k-\omega_s)^2}\left(2\sin(\omega_k-\omega_s)\frac{\Delta t}{2}\right)^2 e^{-i(\omega_k-\omega_s)(l-m)\Delta t}\right)$$

$$= -2\sum_k\sum_{l=1}^{2n-1}\sum_{m=0}^{l-1}\lambda_l\lambda_m\frac{|V_{ks}|^2}{(\omega_k-\omega_s)^2}\left(2\sin(\omega_k-\omega_s)\frac{\Delta t}{2}\right)^2\cos(\{\omega_k-\omega_s\}(l-m)\Delta t)$$

(C.5)

where $|F|^2$ has been neglected since it is $O(|V_{ks}|^4)$. Assuming that $\lambda_j = (-1)^j$,

$$2\operatorname{Re}(F^*G) = -2\sum_k \sum_{l=1}^{2n-1} \sum_{m=0}^{l-1} (-1)^{l+m} \frac{|V_{ks}|^2}{(\omega_k - \omega_s)^2} \left(2\sin(\omega_k - \omega_s)\frac{\Delta t}{2}\right)^2 \cos\left(\{\omega_k - \omega_s\}(l-m)\Delta t\right)$$

$$= -\sum_k |V_{ks}|^2 \tan^2(\omega_k - \omega_s)\frac{\Delta t}{2} \frac{\sin^2(\omega_k - \omega_s)\frac{2n\Delta t}{2}}{\left(\frac{\omega_k - \omega_s}{2}\right)^2} + 2n\sum_k |V_{ks}|^2 \frac{\sin^2(\omega_k - \omega_s)\frac{2n\Delta t}{2}}{\left(\frac{\omega_k - \omega_s}{2}\right)^2} \quad (C.6)$$

Substituting Eq. (C.6) back into Eq. (C.5), one obtains exactly the result derived in Eq. (29), within the context of deterministic pulses. When the variables are allowed to be stochastic, the expression for the survival probability, Eq. (C.5), resembles the one obtained for qubit coherence under similar conditions.

**REFERENCES**

1. Devoret, M. H.; Estive, D.; Urbina, C.; Martinis, J.; Cleland, A.; Clarke, J. Quantum Tunneling in Condensed Media: Amsterdam: North-Holland, 1992.
2. Martinis, J. M.; Devoret, M. H.; Clarke, J. Phys. Rev. B 1987, 35, 4682-4698.
3. Devoret, M. H.; Esteve, D.; Martinis, J. M.; Cleland, A.; Clarke, J. Phys. Rev. B 1987, 36, 58-73.
4. Rice, S. A.; Zhao, M. Optical Control of Molecular Dynamics; John Wiley & Sons, Inc.: New York, 2000.
5. Brumer, P. W.; Shapiro, M. Principles of the Quantum Control of Molecular Processes; John Wiley & Sons, Inc.: New York, 2003.
6. Tannor, D. J.; Rice, S. A. Adv. Chem. Phys. 1988, 70, 441-523.
7. Tannor, D. J.; Rice, S. A. J. Chem. Phys. 1985, 83, 5013-5018.
8. Gordon, R. J.; Rice, S. A. Ann. Rev. Phys. Chem. 1997, 48, 601-641.
9. Shi, S. H.; Woody, A.; Rabitz, H. J. Chem. Phys. 1988, 88, 6870-6883.
10. Chakrabarti, R.; Rabitz, H. Intl. Rev. Phys. Chem. 2007, 26, 671-735.
11. Branderhorst, M. P. A.; Londero, P.; Wasylczyk, P.; Brif, C.; Kosut, R. L.; Rabitz, H.; Walmsley, I. A. Science 2008, 320, 638-643.
12. Shapiro, E. A.; Walmsley, I. A.; Vanov, M. Y. Phys. Rev. Lett. 2007, 98, Art. No. 050501.
13. Rego, L. G. C.; Abuabara, S. G.; Batista, V. S. Quant. Inform. Comput. 2005, 5, 318-334.
14. Rego, L. G. C.; Abuabara, S. G.; Batista, V. S. J. Mod. Opt. 2006, 53, 2519-2532.
15. Rego, L. G. C.; Abuabara, S. G.; Batista, V. S. J. Mod. Opt. 2007, 54, 2617-2627.
16. Frishman, E.; Shapiro, M. Phys. Rev. Lett. 2001, 87.
17. Frishman, E.; Shapiro, M. Phys. Rev. A 2003, 68.
18. Frishman, E.; Shapiro, M. J. Chem. Phys. 2006, 124.
19. Shapiro, E. A.; Walmsley, I. A.; Vanov, M. Y. Phys. Rev. Lett. 2007, 98.
20. Viftrup, S. S.; Kumarappan, V.; Holmegaard, L.; Stapelfeldt, H.; Artamonov, M.; Seideman, T. Phys.Rev. A 2009, 79, 023404.
21. Pelzer, A.; Ramakrishna, S.; Seideman, T. J.Chem.Phys. 2008, 129, 134301.

22. Rego, L. G. C.; Santos, L. F.; Batista, V. S. Ann. Rev. Phys. Chem. 2009, 60, 293-320.
23. Agarwal, G. S.; Scully, M. O.; Walther, H. Phys. Rev. Lett. 2001, 86, 4271-4274.
24. Agarwal, G. S.; Scully, M. O.; Walther, H. Phys. Rev. A 2001, 63, Art. No. 044101.
25. Lidar, D. A.; Chuang, I. L.; Whaley, K. B. Phys. Rev. Lett. 1998, 81, 2594-2597.
26. Viola, L.; Knill, E.; Lloyd, S. Phys. Rev. Lett. 1999, 82, 2417-2421.
27. Viola, L.; Lloyd, S. Phys. Rev. A 1998, 58, 2733-2744.
28. Viola, L.; Santos, L. F. J. Mod. Opt. 2006, 53, 2559-2568.
29. Santos, L. F.; Viola, L. Phys. Rev. A 2005, 72, Art. No. 062303.
30. Santos, L. F.; Viola, L. Phys. Rev. Lett. 2006, 97, Art. No. 150501.
31. Fischer, M. C.; Gutiérrez-Medina, B.; Raizen, M. G. Phys. Rev. Lett. 2001, 87, 040402.
32. Itano, W. M.; Heinzen, D. J.; Bollinger, J. J.; Wineland, D. J. Phys. Rev. A 1990, 41, 2295-2300.
33. Santos, L. F.; Viola, L. New J. Phys. 2008, 10, 083009
34. Kofman, A. G.; Kurizki, G. Nature 2000, 405, 546-550.
35. Misra, B.; Sudarshan, E. C. G. J. Math. Phys. 1977, 18, 756-763.
36. Pascazio, S.; Namiki, M. Phys. Rev. A 1994, 50, 4582-4592.
37. Facchi, P.; Pascazio, S. Phys. Rev. Lett. 2002, 89, Art. No. 080401.
38. Sanchez, C.; Babonneau, F.; Doeuff, S.; Leaustic, A. Ultrastructure Processing of Advanced Ceramics: New York, 1988.
39. Kogan, E. Hait J. Sci. and Eng. A 2008, 5, 174-183.
40. Longhi, S. Phys. Rev. Lett. 2006, 97, 110402-110404.
41. Claude Cohen-Tannoudji; Bernard Diu; Laloe, F. Quantum Mechanics; Wiley Interscience: Paris, 1977.
42. J.J.Sakurai. Modern Quantum Mechanics; Addison Wesley Publishing Company, Inc: Reading, Massachusetts, 1994.
43. Kurnit, N. A.; Hartmann, S. R.; Abella, I. D. Phys. Rev. Lett. 1964, 13, 567-573.
44. Shiokawa, K.; Lidar, D. A. Phys. Rev. A 2004, 69, Art. No. 030302.
45. Slichter, C. P. Principles of Magnetic Resonance; Springer-Verlag: Berlin, 1992.